\documentclass[aps,pre,amsmath,amssymb,twocolumn]{revtex4}
\usepackage{amsmath,amsthm,amssymb}
\usepackage{graphicx}
\usepackage{pstricks}

\newcommand{\be}{\begin{equation}}
\newcommand{\ee}{\end{equation}}
\newcommand{\ba}{\begin{eqnarray}}
\newcommand{\ea}{\end{eqnarray}}
\newcommand{\baa}{\begin{eqnarray}}
\newcommand{\eaa}{\end{eqnarray}}
\newcommand{\ed}{\end{document}}
\newcommand{\lab}[1]{\label{#1}}
\newcommand{\re}[1]{\eqref{#1}}
\newcommand{\ci}[1]{\cite{#1}}

\def\pa{\partial}

\begin{document}

\title {Transparent nonlinear networks}
\author{J.R. Yusupov$^1$, K.K. Sabirov$^{2}$, M. Ehrhardt$^{3}$  and D.U. Matrasulov$^1$}
\affiliation{$^1$Turin Polytechnic University in Tashkent, 17
Niyazov Str., 100095, Tashkent, Uzbekistan\\
$^2$Tashkent University of Information Technologies, 108 Amir
Temur Str., 100200, Tashkent Uzbekistan\\
$^3$Bergische Universit\"at Wuppertal, Gau{\ss}strasse 20, D-42119 Wuppertal,
Germany}

\begin{abstract}
We consider the reflectionless transport of solitons in networks.
The system is modeled in terms of the nonlinear Schr\"odinger equation on metric graphs,
 for which transparent boundary conditions at the branching points are imposed.
 This allows to derive simple constraints, which link equivalent usual Kirchhoff-type
vertex conditions to the transparent ones. Our approach is applied
to a metric star graph. An extension to more complicated graph
topologies is straight forward.
\end{abstract}
\maketitle

\section{Introduction}

Modeling of soliton dynamics in branched structures and networks
is relevant to many practically important tasks arising in optics,
fluid dynamics, condensed matter, biological physics and polymers.
The motivation for such task arises from the fact that highly
efficient transfer of information, charge, heat, spin and optical
signals in the form of solitons requires developing of effective
models providing tools for tunable wave transport in given
low-dimensional materials. Therefore the problem of soliton
transport in branched structures attracted much attention recently
\cite{Zarif,zar2011,Adami,Karim2013,caputo14,DP2015,noja,Our2015,Adami16,Our1,Adami17,dimarecent,Karim2018,KarimNLDE,Our2018,Ourhe,Chsol,Noja19}.

An effective model that can be used for modeling of soliton
dynamics in networks is based on the solution of nonlinear wave
equations on metric graphs. These metric graphs are the set of
bonds (each assigned a length) connected to each other according
to a rule, which is called topology of a graph. Solving the wave
equation in such domain requires imposing boundary conditions
both, at the branching points (vertices) and ends of each branch.
During the past decade, different nonlinear wave equations on
networks have become one of the rapidly developing topics both, in
theoretical and mathematical physics. The early study of the
nonlinear Schr\"odinger equation (NLSE) and soliton dynamics in
networks dates back to the Refs.~\cite{Zarif,zar2011}, where the
integrability of NLSE under certain constraint have been shown.
Later such study has been extended to NLSE on planar graphs
\ci{Our2015}, sine-Gordon \ci{Our1}, nonlinear Dirac
\ci{KarimNLDE} and nonlinear heat \ci{Ourhe} equations. Detailed
study of corresponding stationary problems was presented in the
Refs.~\ci{Karim2013,Adami16,Adami17,Karim2018}.

A very important feature of the wave transport in networks is the transmission of
solitons through the network branching points, which is usually accompanied by
the reflection (backscattering) of a wave at these points.
If reflection dominates compared to transmission  ``resistivity'' of a network with respect to the wave propagation becomes large and this makes such network less
effective for the use of signal transfer.
Therefore, it is quite important from the viewpoint of practical applications,
to reduce such resistivity by providing a minimum of reflection, or by its absence.
This task leads to the
problem of tunable soliton transport in networks, whose ideal result should be
reflectionless transmission of the waves through the branching points of the
structure. For the practical applications in condensed matter, such
transmission implies ballistic transport of charge, spin, heat and other
carriers in low-dimensional branched materials.
The latter is of importance for
the functionalization of low-dimensional materials having branched structure.

Reflectionless transport of solitons in optical fiber networks is
another important problem  for the fiber optics, as many
information-communication devices (e.g., computers, computer
networks, telephones, etc.) use solitons for information (signal)
transfer. Such networks are also used in different optoelectronic
devices. High speed and lossless transfer of information in such
devices require minimum of backscattering or its absence.
Important areas, where the reflectionless or ballistic transport
of optical solitons in networks is required, are molecular
electronics and conducting polymers \ci{Chsol}.

Earlier, possibility for reflectionless transmission of solitons
in networks has been considered in some studies. In particular, it
was found in the Refs.~\cite{Zarif,zar2011} that the transmission
of solitons through the network branching point can be
reflectionless, provided certain constraints are fulfilled. It was
shown also that these constraints provide the integrability of NLS
equation on networks. Later, a similar effect was observed for
other nonlinear PDEs, such as the sine-Gordon equation
Refs.~\cite{Our1} and the nonlinear Dirac equation
Refs.~\cite{KarimNLDE}. In other words, the above studies revealed
a conjecture (at least, for few PDE), which states that if
nonlinear wave equation on a network is integrable, then the
transmission of solitons through the branching points becomes
reflectionless. However, no strict explanation for such conjecture
have been presented in those studies.

In this paper we give a proof of the above conjecture by showing that the
constraints providing such reflectionless transmission and integrability of the
nonlinear Schr\"odinger equation on networks,
link equivalent usual Kirchhoff type vertex boundary conditions to the so-called transparent boundary conditions.
These latter conditions are well studied previously in detail in the
Refs.~\cite{Arnold1998,Ehrhardt1999,Ehrhardt2001,Ehrhardt2002,Arnold2003,Jiang2004,Antoine2008,Ehrhardt2008,Sumichrast2009,Antoine2009,Ehrhardt2010,Klein2011,Arnold2012,Feshchenko2013,Antoine2014,Petrov2016}.

The paper is organized as follows. In the next section we briefly
recall the concept of transparent boundary conditions for the
nonlinear Schr\"odinger equation on a line. Section III provides
extension of the concept of transparent boundary conditions to
solitons in networks described by the nonlinear Schr\"odinger
equation on metric graphs and presents some numerical results.
Finally, Section IV presents some concluding remarks.

\section{Transparent boundary conditions for the nonlinear Schr\"odinger equation on a line}

The problem of transparent boundary conditions (TBC) for the linear partial
differential equations (PDE) is well developed topic in mathematical and
theoretical physics (see, e.g.,
\cite{Arnold1998,Ehrhardt1999,Ehrhardt2001,Ehrhardt2002,Arnold2003,Jiang2004,Antoine2008,Ehrhardt2008,Sumichrast2009,Antoine2009,Ehrhardt2010,Klein2011,Arnold2012,Feshchenko2013,Antoine2014,Petrov2016}
for review).
However, despite such progress, for the nonlinear PDE, the topic
is not well established, yet, due to certain complications of the problem in
nonlinear case. One of the effective approaches for the case is considering the
nonlinear term as potential in linear PDE and called ``potential approach''.
Below we briefly recall this approach following the
Refs.~\ci{Antoine,Matthias2008}.

We consider the wave (particle) motion in a 1D domain $(-\infty,\; +\infty)$
described by the following time-dependent nonlinear Schr\"odinger equation:
\be
   i\pa_t\psi+\pa_x^2\psi+\beta|\psi|^2\psi,\label{nlse1}
\ee
with the initial condition
\be
   \psi(x,0)=\psi_0(x).\label{inc1}
\ee
Derivation of transparent boundary conditions for nonlinear case is rather
complicated than that for the linear one.
However, one can use so-called
potential approach, where Eq.~\re{nlse1} can be considered as the linear
Schr\"odinger equation with the potential $V=\beta|\psi|^2$.
Then, one can rewrite Eq.~\re{nlse1} in a ``linear form'' as
\be
   i\pa_t\psi+\pa_x^2\psi+V\psi=0,\label{nlse2}
\ee

Let us denote by $\psi$ the solution of Eq.~\re{nlse2} and by $v$ the new
unknown defined by the relation
\be
  v(x,t) = e^{-i\nu(x,t)}\psi(x,t),\lab{new}
\ee
where
\be
   \nu(x,t)=\int\limits_0^t{V(x,s)ds}. \label{enu}
\ee
For the time- and space-derivative of $\psi$ we have
\be
   i\pa_t\psi=e^{i\nu}(i\pa_t-V)v,\label{der1}
\ee
\be
\pa_x^2\psi=ie^{i\nu}(\pa_x^2v+2i\pa_x\nu\cdot\pa_xv+i\pa_x^2\nu\cdot
v-(\pa_x\nu)^2v).\label{der2}
\ee
It is clear that the function $v$ satisfies the Schr\"odinger
equation
\be
L(x,t,\pa_x,\pa_t)v=i\pa_tv+\pa_x^2v+A\pa_xv+Bv=0,\label{sce}
\ee
where $A=2i\pa_x\nu$ and
$B=i\pa_x^2\nu-(\pa_x\nu)^2$.

Expanding the factorization \re{sce}, we get
\begin{align}
L&=(\pa_x+i\Lambda^-)(\pa_x+i\Lambda^+)\nonumber\\
&=\pa_x^2+i(\Lambda^-+\Lambda^+)\pa_x+i\,Op(\pa_x\lambda^+)-\Lambda^-\Lambda^+,\label{sceex}
\end{align}
where $\Lambda^\pm=\Lambda^\pm(x,t,\pa_t)$ are classical
pseudodifferential operators, $\lambda_{1/2}^\pm$ are the
principal symbols of operators $\Lambda^\pm$ given by
$\lambda_{1/2}^\pm=\mp\sqrt{-\tau}$ and the function $\tau$ is
inhomogeneous of degree 1 and is an element of $S_S^{1/2}$. The
total symbol $\lambda^\pm=\sigma({\Lambda^\pm})$ of $\Lambda^\pm$
admits an asymptotic expansion in inhomogeneous symbols as
\begin{equation}
\lambda^\pm=\sigma({\Lambda^\pm}) \sim
\sum\limits_{j=0}^{+\infty}{\lambda_{1/2-j/2}^\pm},\label{lambd}
\end{equation}

From \re{sceex} we deduce the system of operators
\be
  i(\Lambda^-+\Lambda^+)=A,\label{oper1}
\ee
\be
  i\,Op(\pa_x\lambda^+)-\Lambda^-\Lambda^+=i\pa_t+B,\label{oper2}
\ee
  which yields the following symbolic system of equations:
\be
   i(\lambda^++\lambda^-)=A,\label{syme1}
\ee
\be
  i\pa_x\lambda^+-\sum\limits_{\alpha=0}^{+\infty}{\frac{(-i)^\alpha}{\alpha!}\pa_\tau^\alpha\lambda^-\pa_t^\alpha\lambda^+}=-\tau+B.\label{syme2}
\ee
 If we identify the terms of order $1/2$ in the
Eq.~\re{syme1}, we obtain $\lambda_{1/2}^-=-\lambda_{1/2}^+$. Then
from Eq.~\re{syme2}, we get \be
\lambda_{1/2}^+=\pm\sqrt{-\tau}.\label{elambda} \ee

The Dirichlet-Neumann operator corresponds to the choice
$\lambda_{1/2}^+=-\sqrt{-\tau}$.
From the factorization~\re{sce} we have the
following transparent boundary condition applied to the unknown wave function $v$
\begin{align}
(-\pa_x+i\Lambda^+)v(0,t)&=0,\label{tbc0}\\
(\pa_x+i\Lambda^+)v(L,t)&=0.\label{tbcL}
\end{align}
Then using Eq.~\re{new} the formal transparent boundary conditions for $\psi$ at
$x=0$ and $x=L$ can be written as \cite{Antoine}
\begin{align}
-&\pa_x\psi(0,t)+e^{-i\frac{\pi}{4}}e^{i\nu(0,t)}\pa_t^{1/2}\left(e^{-i\nu(0,t)}\psi(0,t)\right)=0,\label{nlabc1}\\
&\pa_x\psi(L,t)+e^{-i\frac{\pi}{4}}e^{i\nu(L,t)}\pa_t^{1/2}\left(e^{-i\nu(L,t)}\psi(L,t)\right)=0,\label{nlabc2}
\end{align}
where
\be
\pa_t^{1/2}f(t)
=\frac{1}{\sqrt{\pi}}\pa_t\int_0^t\frac{f(s)}{\sqrt{t-s}}\,ds.\label{frder1}
\ee

Formally, Eqs.~\re{nlabc1} and \re{nlabc2} are similar to those for the linear case.
We remark that a detailed treatment of Eqs.~\re{nlse1}, \re{inc1}, \re{nlabc1} and \re{nlabc2} can
be found in the Refs.~\cite{Antoine,Matthias2008}, where the discretization
scheme and the numerical method for solving of this problem are also presented.
We note that the boundary conditions \re{nlabc1} and \re{nlabc2} are true both,
for focusing ($\beta > 0$) and defocusing ($\beta <0$) cases.
In the next Section III we will modify these boundary conditions for the nonlinear
Schr\"odinger equation on metric graphs.

\section{Transparent boundary conditions for NLSE on metric graphs}
Soliton dynamics in networks is one of the rapidly evolving topics during past
decade.
The early treatment of the problems dates back to the Ref.~\ci{Zarif}, where
soliton solutions of the nonlinear Schr\"odinger equation on metric graphs was
obtained and integrability of the problem under certain constraints was shown
by proving the existence of infinite number of conserving quantities.
An interesting feature found in \ci{Zarif} was the fact that for integrable case,
transmission of solitons through the graph vertices is reflectionless, i.e.\
there is no backscattering of solitons at the graph branching point.
An explanation of such effect was given in recent papers \ci{dimarecent,DP2019},
where it was strictly shown that if the parameters of the generalized Kirchhoff
boundary conditions on a star graph are related to the parameters of the
nonlinear evolution equations and satisfy a single constraint, then the
nonlinear evolution equation on the star graph can be reduced to the
homogeneous equation on the infinite line.
Here we provide more strict proof of this conjecture, by showing that
vertex boundary conditions in the form of
weight continuity and generalized Kirchhoff rules become equivalent to
transparent boundary conditions,
if the parameters of the problem fulfill the
integrability condition given in the form of the sum rule.
To do this, we will apply the above method for imposing transparent boundary
conditions to the NLSE on
metric graphs.
Before doing this, let us briefly recall the treatment of the NLS
equation on metric graphs following Ref.~\ci{Zarif}.

Before this was done for quantum graphs described by the linear Schr\"odinger
equation on metric graphs.
We consider the star graph with three bonds $B_j$
(see, Fig.~1) for which a coordinate $x_j$ is assigned.
Choosing the origin of coordinates at the vertex,
0  for bond $B_1$ we put $x_1\in (-\infty,0]$ and
for $B_{1,2}$ we fix $x_{2,3}\in [0,+\infty)$.
In what follows we use the
notation $\Psi_j(x)$ for $\Psi_j(x_j)$ and $x$ is the coordinate on the bond
$j$ to which the component $\Psi_j$ refers.
The nonlinear Schr\"odinger
equation on each bond of such graph can be written as
\be
  i\,\pa\psi_j+\pa_x^2\psi_j+\beta_j|\psi_j|^2\psi_j=0.\label{nls01}
\ee

\begin{figure}[t!]
\includegraphics[width=70mm]{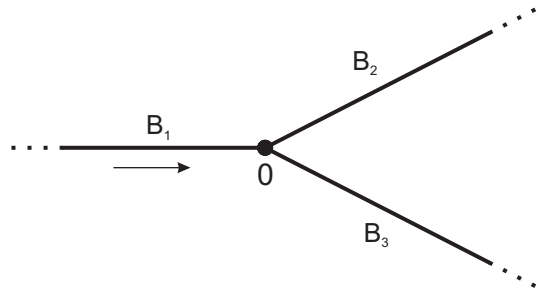}
\caption{Sketch of a star graph with 3 semi-infinite bonds.} \label{pic1}
\end{figure}

Solving this equation \eqref{nls01} requires imposing initial conditions and boundary
conditions at the branching point.
The latter can be derived from the
fundamental physical laws, such as energy and norm conservation, which are
given as
\be
  \frac{dN}{dt} = \frac{dE}{dt} =0, \lab{conserv}
\ee where
\begin{equation*}
N(t) =\int\limits_{-\infty}^{0}|\psi_1|^2\,dx
+\int\limits_{0}^{\infty}|\psi_2|^2\,dx +\int\limits_{0}^{\infty}|\psi_3|^2\,dx
\end{equation*}
and
\begin{equation*}
E =E_1+E_2+E_3,
\end{equation*}
with
\begin{equation*}
E_k =\int\limits_{B_k}\left[\left|\frac{\pa \psi_k}{\pa x}\right|^2
-\frac{\beta_k}{2}|\psi_k|^4\right]\,dx
\end{equation*}
The conservation laws Eq.~\re{conserv} lead to the following vertex conditions
\ci{Zarif}
\be
   \alpha_1\psi_1(0)=\alpha_2\psi_2(0)=\alpha_3\psi_3(0) \lab{vbc1}
\ee
and generalized Kirchhoff rules
\be
 \frac{1}{\alpha_1}\frac{\pa\psi_1}{\pa x}|_{x=0}
=\frac{1}{\alpha_2}\frac{\pa\psi_2}{\pa
x}|_{x=0}+\frac{1}{\alpha_3}\frac{\pa\psi_3}{\pa x}|_{x=0},
\lab{vbc2}
\ee where $\alpha_j$ are nonzero real constants.
The asymptotic conditions for Eq.~\re{nls01} are imposed as
\be
  \psi_j|_{|x|\to +\infty} \to 0. \lab{asymp}
\ee

The single soliton solutions of Eq.~\re{nls01} fulfilling the vertex boundary
conditions \re{vbc1}, \re{vbc2} and the asymptotic condition, \re{asymp} can be
written as \ci{Zarif}
\be
  \psi_j(x,t) = a\sqrt{\frac{2}{\beta_j}}\frac{\exp\bigl[i\frac{vx}{2}-
i(\frac{v^2}{4}-a^2)t\bigr]}{\cosh[a(x-l -vt)]},
\ee
where the parameters $\beta_j$ fulfill the sum rule
\be
  \frac{1}{\beta_1}=
  \frac{1}{\beta_2}+\frac{1}{\beta_3}.\lab{sumrule}
\ee
Here $v$, $l$ and $a$
are bond-independent parameters characterizing velocity, initial center of mass
and amplitude of a soliton, respectively.

\begin{figure}[th!]
\includegraphics[width=84mm]{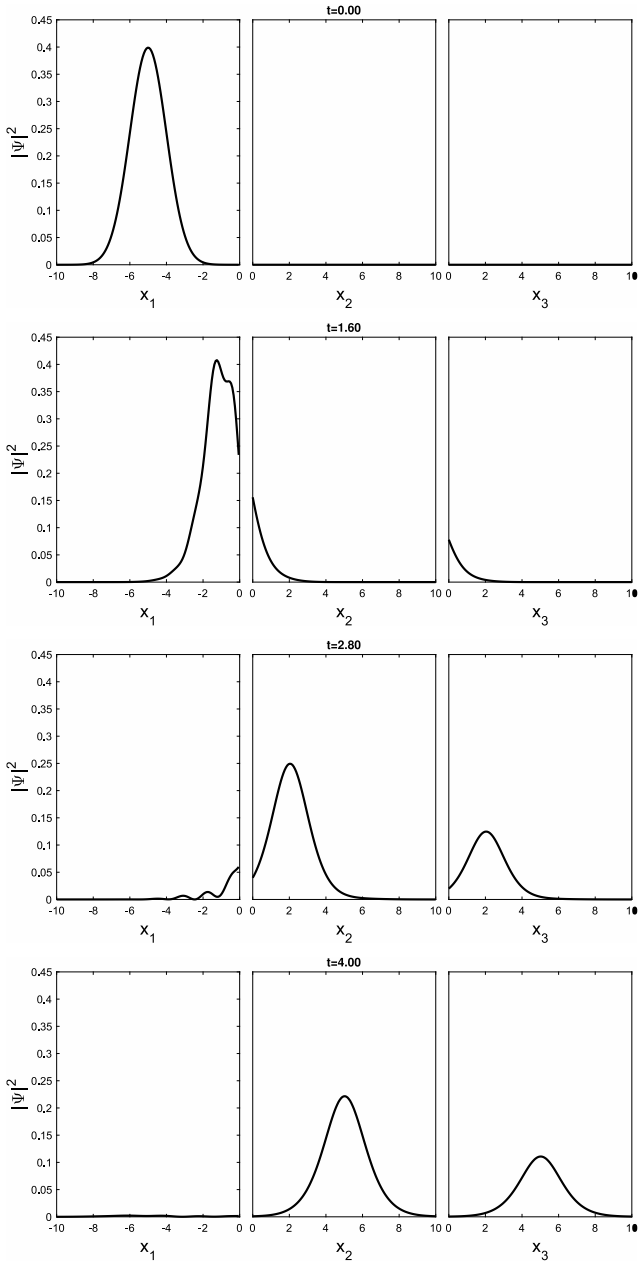}
\caption{The profile of the wave function plotted at different time moments for
the regime when the sum rule is fulfilled (no reflection is occurred):
$\alpha_1=\sqrt{\beta_1}=1/\sqrt{1/2+1/4}$,
$\alpha_2=\sqrt{\beta_2}=\sqrt{2}$
and $\alpha_3=\sqrt{\beta_3}=\sqrt{4}$.
Each column number (from the left to
the right) corresponds to a bond number.} \label{fig2}
\end{figure}

Eq.~\re{sumrule} presents the conditions for integrability of the problem
given by Eqs.~\re{nls01}, \re{vbc1}, \re{vbc2} and \re{asymp},
i.e.\ the integrability of the
nonlinear Schr\"odinger equation on a metric star graph presented in Fig.~1.
It was shown in \ci{Zarif} that under the constraint \re{sumrule}
the problem has an infinite number of constants of motion.
Below we show an additional consequence
following from Eq.~\re{sumrule}, which can be formulated as follows:
If the parameters $\beta_j$ in Eq.~\re{nls01} fulfill the condition (constraint)
\re{sumrule}, then the vertex boundary conditions \re{vbc1} and \re{vbc2}
become equivalent to transparent boundary conditions at the point 0.

Without loss of generality of the approach, we can assume that
$\alpha_j=\sqrt{\beta_j}$.
To impose transparent boundary conditions for
the NLS equation on metric graph shown in Fig.~1,
we split the whole domain (graph) into two
domains called ``interior'' ($-\infty< x < 0$) and ``exterior'' ($0 < x
<\infty$) ones (see, e.g.,
Refs.~\cite{Ehrhardt1999,Ehrhardt2001,Ehrhardt2002,Arnold2003,Jambul} for
details).
Correspondingly, we have interior and exterior problems. The interior
problem is given on $B_1$ by the equations
\begin{align*}
i\,\pa\psi_1+\pa_x^2\psi_1+\beta_1|\psi_1|^2\psi_1&=0,\quad x<0,t>0\\
\psi_1|_{t=0}&=\Psi^I(x),\\
\pa_x\psi_1|_{x=0}&=\left(T_+\psi_1\right)|_{x=0}.
\end{align*}
The exterior problems for $B_{2,3}$ can be written as
\begin{align*}
i\,\pa\psi_{2,3}+\pa_x^2\psi_{2,3}+\beta_{2,3}|\psi_{2,3}|^2\psi_{2,3}&=0,\\
\psi_{2,3}|_{t=0}&=0,\\
\psi_{2,3}|_{x=0}=\Phi_{2,3}(t),\,\Phi_{2,3}(0)&=0,\\
\left(T_+\Phi_{2,3}\right)|_{x=0}&=\pa_x\psi_{2,3}|_{x=0}.
\end{align*}
We rewrite the NLSE of exterior problems for $B_{2,3}$ as
\be
   i\,\pa\psi_{2,3}+\pa_x^2\psi_{2,3}+V_{2,3}\psi_{2,3}=0,\label{nlse3}
\ee
with the potentials $V_{2,3}=\beta_{2,3}|\psi_{2,3}|^2$.
Furthermore, we introduce the new functions $v_{2,3}$ given as
\be
v_{2,3}(x,t)=e^{-i\nu_{2,3}(x,t)}\psi_{2,3}(x,t),\label{newf23}
\ee
where
\be
   \nu_{2,3}(x,t)=\underset{0}{\overset{t}{\int}}V_{2,3}(x,s)\,ds.\label{enu23}
\ee
Then from the factorization in Eq.~\eqref{sceex} we have the
following transparent boundary conditions for the wave functions $v_{2,3}$:
\be
\left(-\pa_x+i\Lambda^+\right)v_{2,3}(0,t)=0.\label{opv23}
\ee
Using Eq.~\eqref{newf23} we can write the formal
transparent boundary conditions for $\psi_{2,3}$ at $x=0$ as
\begin{multline}\label{tbc23}
-\pa_x\psi_{2,3}(0,t)+e^{-i\frac{\pi}{4}}\,e^{i\nu_{2,3}(0,t)}\\
\pa_t^{1/2}\left(e^{-i\nu_{2,3}(0,t)}\psi_{2,3}(0,t)\right)=0.
\end{multline}

Using the vertex boundary condition \re{vbc1} we have
\begin{align}
\pa_x\psi_{2,3}|_{x=0}=&\frac{1}{\sqrt{\pi}}e^{-i\frac{\pi}{4}+i\beta_{2,3}\int\limits_0^t{|\psi_{2,3}(0,s)|^2ds}}\cdot\nonumber\\
&\pa_t\int\limits_0^t{\frac{\psi_{2,3}(0,\tau)e^{-i\beta_{2,3}\int\limits_0^{\tau}{|\psi_{2,3}(0,s)|^2ds}}}{\sqrt{t-\tau}}\,d\tau}\nonumber\\
=&\frac{1}{\sqrt{\pi}}\sqrt{\frac{\beta_1}{\beta_{2,3}}}e^{-i\frac{\pi}{4}+i\beta_{1}\int\limits_0^t{|\psi_{1}(0,s)|^2ds}}\cdot\nonumber\\
&\pa_t\int\limits_0^t{\frac{\psi_{1}(0,\tau)e^{-i\beta_{1}\int\limits_0^{\tau}{|\psi_{1}(0,s)|^2ds}}}{\sqrt{t-\tau}}\,d\tau}.\label{eq4}
\end{align}
From the vertex boundary condition \re{vbc2} and \re{eq4} we get
\begin{align}
\pa_x\psi_1|_{x=0}=&\frac{\sqrt{\beta_1}}{\sqrt{\beta_2}}\pa_x\psi_2|_{x=0}+\frac{\sqrt{\beta_1}}{\sqrt{\beta_3}}\pa_x\psi_3|_{x=0}\nonumber\\
=&\frac{1}{\sqrt{\pi}}\beta_1\left(\frac{1}{\beta_{2}}+\frac{1}{\beta_{3}}\right)e^{-i\frac{\pi}{4}+i\beta_{1}\underset{0}{\overset{t}{\int}}|\psi_{1}(0,s)|^2ds}\cdot\nonumber\\
&\pa_t\underset{0}{\overset{t}{\int}}\frac{\psi_{1}(0,\tau)e^{-i\beta_{1}\underset{0}{\overset{\tau}{\int}}|\psi_{1}(0,s)|^2ds}}{\sqrt{t-\tau}}\,d\tau.\label{tbc01}
\end{align}

It is clear that if the sum rule given by Eq.~\re{sumrule} is fulfilled, i.e.,
$$
  \beta_1\left(\frac{1}{\beta_{2}}+\frac{1}{\beta_{3}}\right) =1,
$$
then the
boundary condition given by Eq.~\re{tbc01} coincides with that in
Eq.~\re{nlabc2}.
Thus fulfilling the sum rule \re{sumrule} implies that vertex
boundary conditions \re{vbc1} and \re{vbc2} become equivalent to transparent
boundary conditions at the graph vertex.
This can be shown by direct numerical
solution of Eq.~\re{nls01} for the boundary conditions \re{vbc1} and \re{vbc2}.
In Fig.~2 the profile of the soliton $|\psi_j|^2$ obtained numerically is
plotted at different time moments for the regime, when the sum rule
\re{sumrule} is fulfilled.
Numerical simulations are performed for the right
traveling Gaussian wave packet given by
\begin{equation*}
    \Psi^I(x)=(2\pi)^{-1/4}\exp(2.5ix-(x+5)^2/4)
\end{equation*}
at four consecutive time steps.

To show that for the case, when the sum rule is broken the transmission of
soliton is accompanied by reflections, we plotted the reflection coefficient,
which is determined as the ratio of the partial norm for the first bond to the
total norm
\begin{equation*}
    R=\frac{N_1}{N_1+N_2+N_3},
\end{equation*}
as a function of $\alpha_1$ for the fixed values of $\alpha_2$ and
$\alpha_3$. It is clear from this plot that the reflection
coefficient becomes zero at the value of $\alpha_1$, which
provides fulfilling of the sum rule \re{sumrule}. This also can be
considered as additional confirmation for becoming equivalent the
vertex boundary conditions in Eqs.~\re{vbc1} and \re{vbc2} to the
transparent ones. It is clear that such conjecture can be derived
for star graph with arbitrary number of bonds. Finally, we note
that the above constraint for transparent boundary conditions
given by Eq.~\re{sumrule} is applicable not only for solitons, but
for arbitrary solutions of the NLS equation on graphs.

\begin{figure}[t!]
\includegraphics[width=90mm]{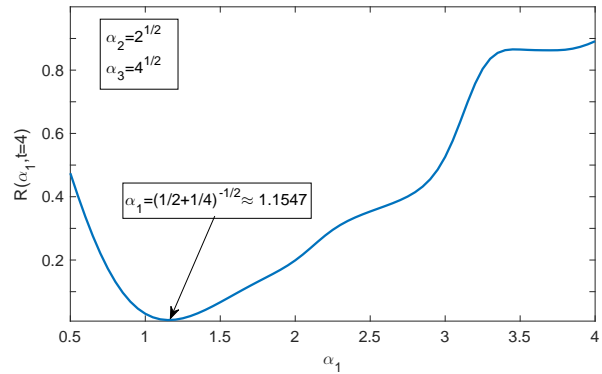}
\caption{Dependence of the vertex reflection coefficient $R$ on the parameter
$\alpha_1$ when time elapses ($t=4$).} \label{alpha}
\end{figure}

\section{Conclusions}
In this paper we studied the problem of reflectionless soliton transport in
network branching points by modeling the soliton dynamics in networks in terms
of the nonlinear Schr\"odinger equation on metric graphs. By combining the
concept of transparent boundary conditions with the Kirchhoff-type boundary
conditions at the vertex, we derived constraints, which make equivalent the
latter to transparent ones.
This gives clear explanation for the previously
observed (see, \ci{Zarif}) conjecture for the absence of soliton
backscattering, when the NLS equation on metric graphs is integrable and
integrability if provided in terms of the above constraint.

Also, solving the problem numerically, we have shown for the star graph a
reflectionless transmission of soliton through the vertex in the case of
fulfilling of the sum rule by the parameters. We note that the approach can be
directly extended to arbitrary graph topologies, which contain any subgraph
connected to two or more outgoing, semi-infinite bonds.
Moreover, we believe
that approach can be extended to other PDE, where similar regime of
reflectionless vertex transmission of sine-Gordon \ci{Our1} and Dirac
\ci{KarimNLDE} solitons have been observed.
We note that the approach used in
this paper can be directly extended to other graphs topologies, such as tree,
loop, triangle, etc., provided the graph consists of finite subgraph and two or
more semiinfinite outgoing bonds.

The above model for reflectionless soliton transport through the network
branching points may have direct and important applications for different
practically important problems of optics, condensed matter and polymers.
Among such applications one can consider optical fiber networks widely used in
computing and communication technologies, where the signal transfer is done in
the form of soliton transport.

\section{Acknowledgements}
This work is partially supported by the grant of the Ministry for Innovation
Development of Uzbekistan (Ref.\ Nr.\ BF2-022).


\end{document}